\documentstyle[aps,prb,multicol,epsf,times]{revtex}
\begin{document}
\draft
\renewcommand{\narrowtext}{\begin{multicols}{2} \global\columnwidth20.5pc}
\renewcommand{\widetext}{\end{multicols} \global\columnwidth42.5pc}
\multicolsep = 8pt plus 4pt minus 3pt
\title{On the Relationship Between the Pseudo- and Superconducting 
Gaps:
  Effects of Residual Pairing Correlations Below $\bbox{T_c}$}
\author{Ioan Kosztin, Qijin Chen, Boldizs\'ar Jank\'o and K.~Levin}
\address{The James Franck Institute, The University of Chicago, 5640
  S.~Ellis Avenue, Chicago IL 60637}
\date{\today}
\maketitle
\begin{abstract}
  The existence of a normal state spectral gap in underdoped cuprates raises
  important questions about the associated superconducting phase. For
  example, how does this pseudogap evolve into its below $T_c$ counterpart?
  In this paper we characterize this unusual superconductor by investigating
  the nature of the ``residual'' pseudogap below $T_c$ and, find that it
  leads to an important distinction between the superconducting excitation
  gap and order parameter.  Our approach is based on a conserving
  diagrammatic BCS Bose-Einstein crossover theory which yields the precise
  BCS result in weak coupling at any $T<T_c$ and reproduces Leggett's
  results in the $T=0$ limit.  We explore the resulting experimental
  implications.
\end{abstract}
\pacs{PACS numbers: 
74.20.-z,
74.25.-q,
74.62.-c,
74.72.-h \hfill [{\sf cond-mat/9805065}]}

\narrowtext

Pseudogap properties, associated with the unusual normal state of the
underdoped high temperature superconductors, have received considerable
attention in the literature.  From an experimental perspective the
relationship (if any) between the pseudo- and superconducting gaps has not
been unambiguously clarified.  Angle-resolved
photoemission\cite{arpesanl,arpesstanford} (ARPES), and other measurements
\cite{sp_heat,NMR} on the underdoped cuprates indicate that the normal state
excitation or, equivalently, pseudo-gap above $T_c$ evolves smoothly into
the excitation gap at and below $T_c$.  It is unlikely that a fully
developed pseudogap will abruptly disappear as the temperature falls below
$T_c$, but precisely how it connects with the superconducting \textit{order
  parameter} is not obvious.
While there are scenarios in which the pseudogap and superconducting gaps
are believed to be inter-related\cite{Emery,Lee} here we point out a
\textit{quantitative} relationship for one widely discussed scenario and
characterize key features of the resulting unusual superconducting state. In
this way we suggest experimental tests which may distinguish one pseudogap
model from another.

A large class of pseudogap scenarios for the underdoped cuprates associate
this phase with some form of precursor
superconductivity\cite{precursor_review}, often in the context of a (normal)
state intermediate between that of the free fermions of the BCS and bound
pairs of the Bose-Einstein regimes \cite{crossover_review}. These BCS
Bose-Einstein crossover theories were originally formulated by
Leggett\cite{Leggett2} to address the nature of the superconducting ground
state.  There has been considerable attention paid to this approach
\cite{NSR,randeria89-981,randeria92-2001,sa93-3202,Haussmann,SereneNSR,%
tchernyshyov97-3372},
primarily at and above $T_c$. Our goal here is to establish a crossover
theory in the regime $0\le T\le T_c$, which is consistent with three
important criteria. These involve \textit{simultaneously} (i) satisfying the
law of particle conservation, (ii) establishing consistency with the precise
BCS result in weak coupling, and (iii) establishing consistency with the
formulation of Ref.~\onlinecite{Leggett2} for the ground state. Of these
three criteria, the first\cite{NSR,SereneNSR} and second, as well as
third\cite{Haussmann,tchernyshyov97-3372} have not necessarily been
satisfied in earlier work.  In this process we determine the counterpart to
the pseudogap below $T_c$, and its experimental implications.

In this paper we build on previous work\cite{Janko,Maly,Qijin}, based on a
particular diagrammatic version of these crossover theories. For
definiteness, we take a simple model of 3D fermions which interact via a
short range, separable pairing interaction with $s$-wave symmetry
$V_{\bbox{k,k'}} = -|g|\,\varphi_{\bbox{k}}\,\varphi_{\bbox{k'}}$.
It should be stressed that we have previously demonstrated that our results
(above $T_c$) for this isotropic model remain qualitatively similar when
applied to a quasi-2D lattice model with attractive $d$-wave
interactions\cite{Qijin}. While, the latter is more suitable for describing
the superconducting state of the cuprates, the general physics we discuss
here is presented more clearly, without the complexity of $d$-wave pairing.
Our diagrammatic scheme is based on the ``pairing approximation'' of
Kadanoff and Martin \cite{kadanoff61}, subsequently extended by
Patton\cite{Patton}, which will be shown below to satisfy the three criteria
discussed above.  Following these references, one arrives at the following
complete set of equations

\begin{mathletters}
  \label{eq:pa}
  \begin{eqnarray}
    \label{eq:pa1}
    \Sigma(K) &=&  G_o^{-1}(K) - G^{-1}(K) \nonumber\\
    &=& \sum_Q t(Q)\,G_o(Q-K)\,\varphi^2_{\bbox{k}-
\bbox{q}/2}\;,\\[1ex]
    \label{eq:pa2}
    g &=& [1+g\,\chi(Q)]\,t(Q)\;,\\[1.5ex]
    \label{eq:pa3}
    \chi(Q) &=& \sum_K G(K)\,G_o(Q-K)\,
    \varphi^2_{\bbox{k}-\bbox{q}/2}\;,\\[1ex] 
    \label{eq:pa4}
    n &=& 2\,\sum_K G(K)\;,
  \end{eqnarray}
\end{mathletters}
which self-consistently determine both the Green's function $G(K)$ and the
T-matrix $t(Q)$. Equation (\ref{eq:pa4}) is associated with particle
conservation, criterion (i).
For brevity, in Eqs.~(\ref{eq:pa}) we have used a four-momentum notation
$K\equiv(\bbox{k};i\omega)$ and $Q\equiv(\bbox{q};i\Omega)$, where
$\omega/\Omega$ are odd/even Matsubara frequencies.  The bare Green's
function is given by $G_o(K)=(i\,\omega - \xi_{\text{\bf k}})^{-1}$, with
$\xi_{\text{\bf k}} = \bbox{k}^2/2m - \mu$ and $n$ is the particle number
density. Here $\Sigma(K)$ is the self-energy and $\chi(Q)$ the pair
susceptibility.  In the weak coupling limit, Eqs.~(\ref{eq:pa}) can be
regarded as a T-matrix formulation of the generalized BCS theory with

\begin{equation}
  \label{eq:t-sc}
  t_{sc}(Q) \;=\; -\frac{\left|\Delta_{sc}\right|^2}{T}\,\delta(Q)\;,
\end{equation}
where 
$\Delta_{sc}$ is the superconducting order parameter, and the Dirac-delta
function guarantees the factorization of the two-particle correlation
function in a manner consistent with off-diagonal long range order. From
Eq.~(\ref{eq:pa1}), the corresponding BCS self-energy is given by

\begin{equation}
  \label{eq:sigma-sc}
  \Sigma_{sc}(K) \;=\; 
\frac{\left|\Delta_{sc}\right|^2\,\varphi_{\bbox{k}}^2}{i\omega +
    \xi_{\bbox{k}}}\;.
\end{equation}
With (\ref{eq:t-sc}) and (\ref{eq:sigma-sc}), Eqs.~(\ref{eq:pa}) yield the
usual BCS gap equation for $\Delta_{sc}$.
As the coupling strength $g$ is increased, the role of pair fluctuations
(representing the mean square deviation of the pairing field from its
average value $|\Delta_{sc}|$) becomes increasingly important and
additional contributions to the T-matrix need to be appended to
Eqs.~(\ref{eq:t-sc}).  These effects are precisely those needed to describe
pseudogap phenomena above $T_c$.

We write the T-matrix below $T_c$ as

\begin{equation}
  \label{eq:t-sc_pg}
  t(Q) \;=\; t_{sc}(Q)+t_{pg}(Q)\;,
\end{equation}
where the ``singular'' $t_{sc}$, given by (\ref{eq:t-sc}), accounts for the
condensate of Cooper pairs, while the ``regular'' $t_{pg}$ describes pair
fluctuations associated with the pseudogap.
Inserting (\ref{eq:t-sc_pg}) into Eq.~(\ref{eq:pa2}), along with
(\ref{eq:t-sc}) and using the filtering property of the Dirac-delta
function, one obtains 

\begin{equation}
  \label{eq:t-pg}
  t_{pg}(Q) \;=\; \frac{g}{1+g\,\chi(Q)}\;,
\end{equation}
and
\begin{equation}
  \label{eq:gap_eq}
  1+g\,\chi(0) \;=\; 0\;.
\end{equation}
This last equation is the self-consistent gap equation.  Moreover, we see
from the above two equations that the pseudogap component of the T-matrix,
$t_{pg}(Q)$, is highly peaked about the origin, with a divergence at $Q=0$
\cite{thouless}.
The self-energy of Eq.~(\ref {eq:pa1}) may be decomposed into 
two contributions 

\begin{equation}
  \label{eq:sigma-sc-pg}
  \Sigma(K) \;=\; \Sigma_{sc}(K) + \Sigma_{pg}(K)\;.
\end{equation}
In evaluating the pseudogap contribution to the self-energy, detailed
numerical calculations\cite{Maly} show that the main contribution to the $Q$
sum comes from the the small $Q$ region so that $\Sigma_{pg}(K) =
G_o(-K)\,\varphi^2_{\bbox{k}}\sum_Q t_{pg}(Q) + \delta\Sigma(K)$,
where $\delta\Sigma$, in the momentum and frequency range of interest, is
much smaller than the leading BCS-like part, and can be ignored in what
follows\cite{new2}.
Thus, $\Sigma_{pg}$ can be well approximated by the BCS-like form

\begin{equation}
  \label{eq:sigma-pg}
  \Sigma_{pg}(K) \;\approx\;
\frac{\Delta_{pg}^2\,\varphi^2_{\bbox{k}}}{i\,\omega+\xi_{\bbox{k}}}
\;, 
\end{equation}
where the pseudogap amplitude within the superconducting state,
$\Delta_{pg}$, is defined as

\begin{equation}
  \label{eq:pg}
  \Delta^2_{pg} \;=\; -\sum_{Q} t_{pg}(Q)\;.
\end{equation}
Note that, although $|\Delta_{sc}|$ satisfies an equation similar to
(\ref{eq:pg}), there is an important distinction between the two energy
gaps.  $\Delta_{sc}$ is a complex order parameter which represents the mean
value of the pairing field.  By contrast, the pseudogap parameter
$\Delta^2_{pg}$ is a positive definite quantity which describes the
(incoherent) fluctuations of the pairing field about its mean value.

It follows from Eqs.~(\ref{eq:sigma-sc-pg})-(\ref{eq:pg}) that 

\begin{equation}
  \label{eq:sigma-a}
  \Sigma(K) \approx \frac{\Delta^2\,\varphi^2_{\bbox{k}}}{i\omega+
  \xi_{\bbox{k}}}\;, \quad 
\Delta\equiv\sqrt{\left|\Delta_{sc}\right|^2+\Delta_{pg}^2}\;.
\end{equation}

The central results for the superconducting gap equation, number density and
pseudogap below $T_c$ follow from
Eqs.~(\ref{eq:pa},\ref{eq:t-pg},\ref{eq:gap_eq},\ref{eq:pg},\ref{eq:sigma-a})
and can be summarized as

\begin{mathletters}
  \label{eq:gap}
  \begin{eqnarray}
    \label{eq:gap1}
    0 &=& 1+g\sum_{\text{\bf k}} \frac{1- 2\,f\left(E_k\right)}{2\,E_k}\,
    \varphi^2_{\bbox{k}}\;, \\ 
    \label{eq:gap2}
    n &=& \sum_{\text{\bf k}}\left[1-\frac{\xi_k}{E_k} +
      \frac{2\,\xi_k}{E_k}\,f\left(E_k\right)\right] \;,\\
    \label{eq:gap3}
    \Delta^2_{\text{pg}} &=& - \sum_{Q} \frac{g}{1+g\sum_{K}
      G(K)\,G_o(Q-K)\,\varphi^2_{\bbox{k}-\bbox{q}/2}}\;.
  \end{eqnarray}
\end{mathletters}
where $E_{\bbox{k}}\,=\,(\xi_k^2 + \Delta^2\varphi^2_{\bbox{k}})^{1/2}$.
Note that the total excitation gap $\Delta$ and the chemical potential $\mu$
can be obtained from first two Eqs.~(\ref{eq:gap}).  Moreover, while
Eqs.~(\ref{eq:gap1},\ref{eq:gap2}) coincide formally with the corresponding
(weak coupling) BCS equations, here these equations are valid for arbitrary
coupling and any $T\le T_c$. It should be stressed that Equation
(\ref{eq:gap3}), which determines the precise decomposition of $\Delta$ into
$\Delta_{sc}$ and $\Delta_{pg}$, is crucial and contains much of the central
new physics of this paper.

The simplicity of these equations derives directly from the diagrammatic
scheme of Eqs.~(\ref{eq:pa}); alternative schemes
\cite{Haussmann,tchernyshyov97-3372} will not produce this standard form,
nor will they lead to the BCS limit in the weak coupling case. Moreover, as
will be seen below, when $T\rightarrow 0$, the pseudogap $\Delta_{pg}$
vanishes and Eqs.~(\ref{eq:gap1},\ref{eq:gap2}) coincide precisely with
those used by Leggett \cite{Leggett2} in his $T=0$ BCS Bose-Einstein
crossover theory.  Finally, as $T\rightarrow T_c$ from below, the equations
satisfied by $T_c$, $\Delta_{pg}$ and $\mu$ can be seen to be identical to
their counterparts found earlier\cite{Maly} when $T_c$ is approached from
the normal state.

Physically, the pseudogap below $T_c$ can be interpreted as an extra
contribution to the excitation gap, reflecting the fact that at moderate and
large $g$, additional energy is needed to overcome the residual attraction
between excited fermion pairs in order to produce fermionic-like Bogoliubov
quasi-particles. In the bosonic limit, it becomes progressively more
difficult to break up these pairs and the energy $\Delta_{pg}$ increases
accordingly.

\begin{figure}
   \centerline{\epsfxsize=3.5in\epsffile{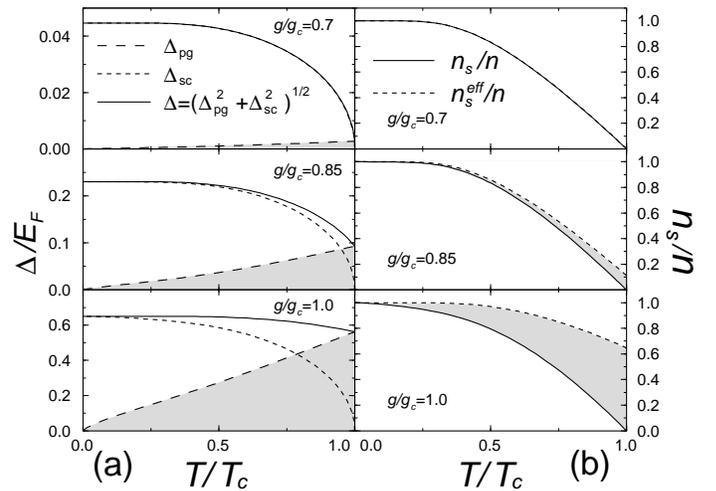}}
   \vspace*{1ex}
 \caption{(a) Temperature dependence of the excitation gap $\Delta$,
   superconducting gap $|\Delta_{sc}|$ and pseudogap $\Delta_{pg}$ for
   coupling strengths $g/g_c = 0.7$, $0.85$ and $1.0$.
   (b) Temperature dependence of the superfluid densities $n_s$ and
   $n_s^{\text{eff}}$ for the same coupling strengths as in (a). The shaded
   regions emphasize pseudogap effects.}
  \label{Fig}
\end{figure}

In Fig.~\ref{Fig}a are plotted $\Delta_{pg}$, $\Delta_{sc}$ and $\Delta$, as
a function of temperature, obtained from a numerical solution of
Eqs.~(\ref{eq:gap}). We choose for illustrative purposes three
representative values for $g/g_c=0.7$, $0.85$ and $1.0$, (all of which lead
to positive chemical potential), corresponding, respectively, to a small,
intermediate and large pseudogap parameter at $T_c$.  Here, for
definiteness, we follow Ref.~\onlinecite{NSR} and take $\varphi_{\bbox{k}} =
(1+k^2/k_o^2)^{-1/2}$, with $k_o=4\,k_F$ and define $g_c=- 4\pi/m\,k_o$ to
represent the critical coupling necessary to form a bound fermion pair in
vacuum.  As can be seen from the figure, with decreasing temperature,
$\Delta_{pg}(T)$ decreases monotonically from its maximum value at $T_c$
until it essentially vanishes \cite{pg_T=0} at $T=0$, while $\Delta_{sc}(T)$
and $\Delta(T)$ both increase monotonically. When approached from slightly
above $T_c$, there will be a slope discontinuity in $\Delta$ at $T_c$,
reflecting the related discontinuity in the order parameter $\Delta_{sc}$;
moreover, as demonstrated in Fig.~\ref{Fig}a, at the higher value of $g$
this total gap is almost temperature independent.

The pseudogap is an important measure of the distinction between the order
parameter, $\Delta_{sc}$, and the excitation gap $\Delta$.  The latter is
the quantity deduced in ARPES measurements. The former must be directly
related to the superfluid density, $n_s$, which is strictly zero at and
above $T_c$, but which presumably also depends in some way on $\Delta$ as
well. Moreover, by studying $n_s$, (which can be obtained via the London
penetration depth), it will be made clear that, in principle, all three gap
parameters can be distinguished experimentally.  The superfluid density may
be expressed in terms of the local (static) electromagnetic response kernel
$K(0)$ \cite{agd}

\begin{equation}
  \label{eq:ns2}
  n_s = \frac{m}{e^2} K(0) = n - \frac{m}{3\,e^2} P_{\alpha\alpha}(0) \;,
\end{equation}
with the current-current correlation function given by 

\begin{eqnarray}
  \label{eq:ns3}
  P_{\alpha\beta}(Q) &=& -2\,e^2 \sum_K
  \lambda_{\alpha}(K,K+Q)\,G(K+Q)\nonumber\\ &&
  \times\,\Lambda_{\beta}(K+Q,K)\,G(K) \;.
\end{eqnarray}
Here the bare vertex $\bbox{\lambda}(K,K+Q) = \frac{1}{m}
(\bbox{k}+\bbox{q}/2)$, while the renormalized vertex $\bbox{\Lambda}$ must
be deduced in a manner consistent with the generalized Ward identity,
applied here for the uniform static case: $Q=(\bbox{q},0)$,
$\bbox{q}\rightarrow 0$ \cite{JRS-sc}. It is convenient to write
$\bbox{\Lambda} = \bbox{\lambda} + \bbox{\delta\Lambda}_{pg}
+\bbox{\delta\Lambda}_{sc}$, where the pseudogap contribution
$\bbox{\delta\Lambda}_{pg}$ to the vertex correction follows from the Ward
identity

\begin{equation}
  \label{eq:dL-pg}
  \bbox{\delta\Lambda}^{pg}(K,K)=\partial\Sigma_{pg}(K)/\partial\bbox{k}\;.
\end{equation}
The particle density $n$, given by Eq.~(\ref{eq:pa4}), after partial
integration can be rewritten as $n=-(2/3)\sum_K
\bbox{k}\cdot\partial{G(K)}/\partial{\bbox{k}}$. Then, as a result of
Dyson's equation, one arrives at the following general expression

\begin{equation}
  \label{eq:ns11}
  n = - \frac{2}{3} \sum_K G^2(K) \left[\frac{\bbox{k}^2}{m} +
    \bbox{k}\cdot\frac{\partial\Sigma_{pg}(K)}{\partial\bbox{k}} +
    \bbox{k}\cdot\frac{\partial\Sigma_{sc}(K)}{\partial\bbox{k}}\right] \;.
\end{equation}
Now, inserting Eqs.~(\ref{eq:ns11}) and (\ref{eq:ns3}) into
Eq.~(\ref{eq:ns2}) one can see that the pseudogap contribution to $n_s$
drops out by virtue of Eq.~(\ref{eq:dL-pg}); we find

\begin{equation}
  \label{eq:n_s-0}
  n_s = \frac{2}{3} \sum_K
  G^2(K)\,\bbox{k}\cdot\left(\bbox{\delta\Lambda}_{sc} -
    \frac{\partial\Sigma_{sc}}{\partial\bbox{k}}\right) \;.
\end{equation}
We emphasize that the cancellation of this pseudogap contribution to the
Meissner effect is solely the result of local charge conservation.

Following the standard prescription for constructing the proper vertex
correction corresponding to the superconducting
self-energy \cite{JRS-sc} one obtains

\begin{equation}
  \label{eq:dL_sc}
  \bbox{\delta\Lambda}_{sc}(K+Q,K) = \Delta_{sc}^2\varphi^2_{\bbox{k}}
  G_o(-K-Q) G_o(K) \bbox{\lambda}(K+Q)\;.
\end{equation}
Inserting Eqs.~(\ref{eq:sigma-sc},\ref{eq:sigma-sc-pg},\ref{eq:dL_sc}) into
Eq.~(\ref{eq:n_s-0}), after calculating the Matsubara sum, one arrives at

\begin{equation}
  \label{eq:n_s}
  n_s = \frac{2}{3} \sum_{\bbox{k}}
  \frac{\Delta_{sc}^2\,\varphi^2_{\bbox{k}}}{E_{\bbox{k}}^2}
  \left[\xi_{\bbox{k}}(3-\varphi_{\bbox{k}}^2)+2\mu\right]\,
  \left[\frac{1-2\,f(E_{\bbox{k}})}{2\,E_{\bbox{k}}} + f'(E_{\bbox{k}})
  \right] \;.
\end{equation}
We may write the superfluid density as $n_s = \Delta^2_{sc}\,F(\Delta)$
[where the form of the function $F$ can be obtained from
Eq.~(\ref{eq:n_s})]. The same quantity corresponding to a BCS superconductor
with effective gap parameter $\Delta$ is given by $n_s^{\text{eff}} =
\Delta^2\,F(\Delta)$, so that, $n_s/n_s^{\text{eff}}=(\Delta_{sc}/\Delta)^2 \le
1$.
In Fig.~\ref{Fig}b we plot the temperature dependence of the normalized
superfluid density $n_s/n$ (solid line) calculated from Eq.~(\ref{eq:n_s})
for the same three representative values $g/g_c$ as above. These curves are
compared (dashed line) with the quantity $n_s^{\text{eff}}/n$, which is a
(BCS-like) function only of the excitation gap. For sufficiently weak
coupling ($g/g_c\lesssim 0.7$) the two curves are indistinguishable. With
increasing $g$ the separation between the two curves become evident,
particularly in the vicinity of $T_c$, whereas at zero temperature there is
no difference since $n_s = n $, independent of the coupling.  This
comparison thus demonstrates how different are these ``pseudogap''
superconductors. The superfluid density reflects most directly the
temperature dependence of $\Delta_{sc}$, \textit{not} the excitation gap.

The existence of residual pairing correlations below $T_c$ will affect
thermodynamic properties as well.  Indeed, upon analysis of data in
underdoped cuprates, Loram \textit{et al\/} \cite{loram94-243} conjectured
that the measured excitation gap squared can be expressed as the sum of the
squares of a pseudogap and superconducting order parameter. This purely {\em
  phenomenological} analysis leads to a similar
decomposition\cite{loram97-1405} of the excitation gap, as in
Eq.~(\ref{eq:sigma-a}) However, in contrast to the present work, these
authors presumed that $\Delta_{pg}$ is temperature independent below $T_c$.

In summary, in this paper we have demonstrated that, if a pseudogap state
arises from pairing correlations (fluctuations) above $T_c$, then these
pairing fluctuations necessarily persist below $T_c$. These pseudogap
systems are unconventional superconductors, in which pair fluctuations are
present all the way down to the lowest temperatures. At $T=0$ these
fluctuations (or $\Delta_{pg}$) vanish.  A key manifestation of the
``superconducting pseudogap'' is in the nature of the excitation gap
($\Delta$), which differs significantly from the superconducting order
parameter $\Delta_{sc}$, as $\Delta^2 = \Delta_{sc}^2+\Delta_{pg}^2$.
At a physical level we view $\Delta_{pg}$ as reflecting an additional energy
associated with the attractive interaction, which must be overcome in order
to create fermionic-like Bogoliubov quasi-particles. In this way, the {\em
  excitations} from the condensate in a BCS Bose-Einstein crossover theory
can be viewed as intermediate between the (free) fermionic Bogoliubov
quasi-particles of the BCS limit and the (bound) bosonic pairs in the
Bose-Einstein regime. It should be stressed that our previous work on
$d$-wave superconductors\cite{Qijin} reinforces the claim that the physics
presented here for the $s$-wave case, is not qualitatively sensitive to the
symmetry of the pairing interaction.

Experimentally, verification of this pseudogap scenario (for the underdoped
cuprate superconductors) involves establishing the relation between $\Delta$
and $\Delta_{sc}$.  Measurements of $n_s$ and $\Delta$ separately are
possible (through penetration depth and ARPES experiments).
Even more promising may be tunneling spectroscopy measurements of high $T_c$
superconductor-insulator-superconductor junctions in which the Josephson and
quasiparticle current data can be simultaneously used to extract $\Delta$
and $\Delta_{sc}$.

We gratefully acknowledge useful discussions with A.~Abrikosov, G.~Mazenko,
M.~Norman and A.~Zawadowski.
This work was supported in part by the Science and Technology Center for
Superconductivity founded by the National Science Foundation under award
No.~DMR91-20000.



\widetext

\begin{references}
\bibitem{new2}
$\delta\Sigma$ can be safely ignored for the purposes of calculating physical
  quantities which involve only integrals over the self energy (e.g., $T_c$,
  $\Delta_{sc}$, etc.) but it must be retained for calculating spectral
  functions, and densities of states, etc. Moreover, by including
  $\delta\Sigma$ in the self energy $\Sigma_{pg}$, its simple BCS form is
  spoiled: the Bogoliubov quasiparticles acquire a finite lifetime.
\bibitem{pg_T=0} 
Even if one takes into account the neglected
  $\delta\Sigma$ term, $\Delta_{pg}$ still remains negligibly small
  compared to the total excitation gap $\Delta$ at $T=0$.
\end{references}
\end{document}